\documentclass[12pt]{article}

\usepackage{scicite}

\usepackage{times}

\newenvironment{sciabstract}{%
\begin{quote} \bf}
{\end{quote}}

\newcommand{\cfmaskcasesny}{$71000$}
\newcommand{\cfmaskcasesciny}{($95\%$ $CI$ $64000$ to  $78000$)}
\newcommand{\cfmaskcasescon}{$29000$}
\newcommand{\cfmaskcasescicon}{($95\%$ $CI$ $27000$ to  $31000$)}
\newcommand{\cfmaskcasesmass}{$68000$}
\newcommand{\cfmaskcasescimass}{($95\%$ $CI$ $65000$ to  $70000$)}

\newcommand{\cfmaskcasesnyc}{$17900$}
\newcommand{\cfmaskcasescinyc}{($95\%$ $CI$ $15800$ to  $20000$)}

\newcommand{\cfmaskcases}{$170000$}
\newcommand{\cfmaskcasesci}{($95\%$ $CI$ $160000$ to  $180000$)}

\newcommand{\serialinterval}{$7.5$}
\newcommand{\serialintervalci}{($95\%$ CI $5.3$ to $19$$5.3$)}
\newcommand{\rnought}{$2.1$}

\newcommand{\rnoughtma}{$1.9$}

\newcommand{\ronema}{$1.1$}

\newcommand{\rtwoma}{$0.66$}

\newcommand{\rone}{$1.6$}

\newcommand{\rtwo}{$0.72$}

\newcommand{\rtwodate}{March 30}
\newcommand{\rthree}{$0.44$}

\newcommand{\rthreedate}{April 14}
\newcommand{\virus}{\mbox{SARS-CoV-2}}
\newcommand{\disease}{\mbox{COVID-19}}

\usepackage{url}
\usepackage{graphicx}
\title{Masks and COVID-19: a causal framework for imputing value to public-health interventions}

\author{Andres Babino, Marcelo O. Magnasco\\ \textit{Laboratory of Integrative Neuroscience, Rockefeller University}}
\author
{Andres Babino,$^{1\ast}$ Marcelo O. Magnasco,$^{1}$\\
\\
\normalsize{$^{1}$Laboratory of Integrative Neuroscience, Rockefeller University,}\\
\\
\normalsize{$^\ast$Corresponding author. E-mail:  ababino@rockefeller.edu.}
}

\date{}

\begin{document}

\baselineskip24pt

\maketitle

\begin{sciabstract}
  During the \disease\ pandemic, the scientific community developed predictive models to evaluate potential governmental interventions. However,  the analysis of the effects these interventions had is less advanced.
  Here, we propose a data-driven framework to assess these effects retrospectively.
  We use a regularized regression to find a parsimonious model that fits the data with the least changes in the $R_t$ parameter.
  Then, we postulate each jump in $R_t$ as the effect of an intervention.
  Following the do-operator prescriptions, we simulate the counterfactual case by forcing $R_t$ to stay at the pre-jump value.
  We then attribute a value to the intervention from the difference between true evolution and simulated counterfactual.
  We show that the recommendation to use facemasks for all activities would reduce the number of cases by \cfmaskcases\ \cfmaskcasesci\ in Connecticut, Massachusetts, and New York State.
  The framework presented here might be used in any case where cause and effects are sparse in time.
\end{sciabstract}

The burst of the \disease\ pandemic forced governments around the globe to take health care interventions.
One that caused significant controversies is the use of masks among the general public \cite{Prather2020,Morawska2020,Feng2020}.
Initially, it was assumed that the primary mode of transmission of \virus\ was through coughing or contact with surfaces.
Extensive shortages of personal protective equipment for health workers led to an initial recommendation for the general population not to wear masks.
While no extant studies show surgical masks reduce transmission of \virus\ in humans, surgical masks do reduce viral shedding of other coronaviruses \cite{Leung2020}, and transmission of \virus\ in animal models \cite{chansurgical}.
Also, cloth masks might filter \virus\cite{Konda2020}.

The  Centers for Disease Control and Prevention (CDC) changed its guidelines on April 3, 2020, and recommended the widespread use of masks  \cite{CDCmasks}.
According to the CDC, the rationale behind its policy change was the increase in evidence that asymptomatic and presymptomatic people are infectious \cite{Rothe2020,Zou2020,Pan2020,Bai2020,Kimball2020,Wei2020} and that there
are many undetected cases \cite{Li2020science}.
On June 5, the World Health Organization (WHO) changed its guidelines and recommended governments to encourage the general public to wear masks in specific situations, like grocery stores \cite{world2020advice}.
However, WHO specifies: ``At present, there is no direct evidence (from studies on COVID19 and in healthy people in the community) on the effectiveness of universal masking of healthy people in the community to prevent infection with respiratory viruses, including COVID-19'' \cite{world2020advice}.
Meanwhile, according to The European Centre for Disease Prevention and Control (ECDC), there is no evidence that masks can prevent people from being infected \cite{ECDCmasks},
and they suggest the use of masks may increase the risk of infection.
The reason for the WHO and ECDC statements is that there is no work on the widespread use of masks by the general public as a non-pharmaceutical intervention (NPI).
In this paper, we aim to fill this gap by testing the hypothesis that the policy change regarding masks by the CDC (and local governments) decreased the number of positive cases in the states of Connecticut (CT), Massachusetts (MA), New York (NY), Rhode Island (RI), and Virginia (VA).

To assess causality, we need to evaluate both branches of an intervention \cite{pearl2009causality}: one in which the intervention did happen, and one in which it did not.
The gold standard to do so is the double-blind randomized control trial (RCT) paradigm.
Although possible \cite{Haushofer2020}, RTCs are not the norm in public health epidemiological intervention.
Even if implemented, there is no such thing as a ``placebo arm'' for travel restrictions or a double-blind school closure.
Since a placebo or double-blind trials are not possible, there is an indirect causal path between the treatment and the outcome \cite{KennedyShaffer2020,Kahn2018}.
For example, people in zip codes with open schools might be more careful with their hygiene because they know that they are at a higher risk than people in zip codes where the schools are closed.
When the second branch of the intervention did not happen, it is called a counterfactual (``contrary to the facts'').
One option to measure the direct effect of an intervention, and the one that we use in this work, is to estimate or simulate the counterfactual branch \cite{NBERw26906,Tian2020}.

Here, we present a framework to analyze data from the \disease\  epidemic that can simulate counterfactual scenarios in which specific NPI did not occur.
In this framework, we use the odds of a positive test as the dependent variable, rather than the number of positive tests \cite{Jonas2020} or deaths \cite{Flaxman2020}.
We motivate a linear equation for the evolution of this dependent variable using the Susceptible-Infected-Recovered (SIR) model \cite{kermack1927contribution,Hethcotet2016}.
Finally, we carry out a LASSO \cite{Tibshirani1996} regression to fit the data, to obtain a piecewise-linear fit to the logarithm of the odds with the smallest number of breaks (see the supplementary material).
This regression finds the times when interventions started, allowing us to simulate alternative scenarios where these interventions did not happen and assess their net impact.

The daily number of cases and tests are highly variable (Fig. \ref{fig:raw_data}).
To reduce this variability, we compute the \textit{log-odds}, the logarithm of the number of positive tests over the number of negative ones.
Doing this, reveals a piecewise linear pattern that we fit using the LASSO regression (Fig. \ref{fig:odds_fit}).
These regressions show three breakpoints in the \textit{log-odds} in NY, two in CT, MA, MI, and RI, and one in VA.
We should stress the fact that these breaks are not an input of the user.
On the contrary, this is the result of applying the LASSO regularization.
These changepoints happen after different NPIs.
The first change in CT and MA, and the first and second in NY, are due to mobility restrictions (school closure, ban mass gatherings, restriction the non-essential workforce, and stay-at-home orders).
In these states, the last break happens after the CDC changed its guidelines regarding masks.
In MI, RI, and VA, the stay at home orders and the CDC recommendation to wear masks happened closer in time, making it hard to disentangle their effects.
However, the MI government only enforced the use of masks in closed areas (like grocery stores), and the VA government never recommended the use of masks.
We assume that the lack of local orders correlates negatively with local compliance with the CDC guidelines, and that explains why we do not see the masking effect on MI and VA.

We use the slopes of the regression to compute the $R_t$, the instantaneous reproduction number \cite{Bettencourt2008}.
This parameter expresses the average number of people infected from one positive case on day $t$.
In Fig. \ref{fig:rt}, we show $R_t$ as a function of time. 
The NY plot, in Fig. \ref{fig:rt}, shows $R_t$ dropping down from \rnought\ to \rone\ and then to \rtwo\ on \rtwodate, $8$ days after the closure of all nonessential business.
Taking them together, this translates to a reduction by 65 \% on $R_t$ due to mobility restrictions.
There is a third drop from \rtwo\ to \rthree\ on \rthreedate, 11 days after the CDC changed their guidelines and recommended to wear masks, and two days after NY enforced the use of masks for public employees.
In CT, the stay-at-home orders reduced the value of $R_t$ by $51 \%$.
Moreover, after the new CDC recommendation on masks, it dropped by $40 \%$.
Remarkably, in MA, after the stay at home order the $R_t$ value dropped from \rnoughtma\ to \ronema, still above $1$.
Only after the recommendation of wearing masks it fell to \rtwoma, below 1.
As we already mentioned, we do not see the effect of masks in MI or VA, and we attribute this to the lack of local compliance.
In MI, the government only enforced the use of masks in enclosed areas, and the VA government never ruled on the use of masks.
The data from RI is harder to interpret because stay-at-home orders and masks guidelines happened close in time, and data from before April are unreliable (with less than 500 tests a day).
Nonetheless, there is an effect from the CDC guidelines and from local governments making masks mandatory.

Finally, one advantage of the sparsifying framework is that we can simulate counterfactual scenarios by removing a breakpoint.
Then, the regressed line would have continued at the previous slope.
Take the case of the public wearing masks.
From the fit shown in Fig. \ref{fig:odds_cf}, we observe that on \rthreedate\ in NY, $R_t$ changed from \rtwo\ to \rthree.
We interpret that the counterfactual to this intervention is that if the public had not used masks, $R_t$ would have stayed at \rtwo, or in causal inference jargon \textit{do(no masks)}.
Fig. \ref{fig:odds_cf} (green line) shows that removing the intervention would have resulted in a much more drawn-out dwindling of the case curve.
Now, we can use the counterfactual odds to calculate the counterfactual number of positive and compare it with the actual number of positive cases
Doing this yields that, between Apr. 14 and May 15, wearing masks had the effect of decreasing the number of infections by \cfmaskcasesny\ cases \cfmaskcasesciny, in NY.
Similarly, the use of masks reduced the number of positive cases by \cfmaskcasesmass\ cases \cfmaskcasescimass\ in MA between Apr. 13 and May 19, and by \cfmaskcasescon\ cases \cfmaskcasescicon\ in CT from Apr. 14 to May 17.

In conclusion, we found that masks reduced the spread of the virus in CT, MA, and NY.
In those states, our calculations showed that the intervention reduced the $R_t$ by 40\%, and we estimate that masking prevented \cfmaskcases\ cases \cfmaskcasesci\ from the moment they were adopted until the end of the stay-at-home orders.
These results are consistent with recently published results by Mitze and colleagues that found the same effect in Germany \cite{mitze2020face}.
Also, we estimate that in New York City alone, masks reduced the number of cases by \cfmaskcasesnyc\ \cfmaskcasescinyc\ between April 17 and May 9.
This number is below the one estimated by Zhang and colleagues who estimated a reduction in 66000 cases during the same period in New York City \cite{Zhang2020}.
We believe the discrepancy can be accounted for by noting that Zhang and colleagues did not consider the increase in testing during the stay-at-home order in,
leading to a higher difference after the masking order at which point the testing was more stable.

The framework that we presented is data-driven, and therefore it relies on only a handful of hypotheses as  compared to other methods
For example, the counterfactual analysis relies on one hypothesis: the \textit{log-odds} are piecewise linear (see Eq. 6 in the supplementary material)--without the need to assume any of the hypotheses of the SIR model.
Based on the goodness of fit, we are confident that this hypothesis holds for the datasets presented here (see Fig. 2 and the supplementary material).
Also, to put the framework to the test, we apply the method to synthetic data, and we found that it was able to find the corresponding breakpoints and slopes (see the supplementary text).
Nevertheless, when there are no reliable figures on negative test results, our framework fails to fit the data (see the supplementary material).
More detailed data will be necessary to build better models in the future.
Ideally, the data would be organized in a case-by-case fashion and it would contain information on the sample criterion.
Fundamentally, this information should also be available also for negative tests.

Overall, we found evidence that masks reduce the spread of the \virus\  and prevent new infections.
We hope that our findings will persuade local authorities and intergovernmental institutions to strongly recommend the use of masks to prevent the spread of \virus.
We arrived at this conclusion by merging two different traditions: causal inference and regularized regression.
We believe that the union of these techniques will be fruitful in other contexts where the causes and effects are sparse in time.

\section*{Acknowledgments}
We thank Will Meierjurgen Farr, Gustavo Stolovitzky, and Joel E. Cohen for their comments and suggestions.
\textbf{Authors contributions:} AB performed the data curation, the formal analysis and the software development.
MOM supervised the project and provided resources.
All authors contributed to the conceptualization and the preparation the original draft.
\textbf{Competing interests:} The authors declare no competing interests.
\textbf{Data and materials availability: } all the data and codes to reproduce our analysis are publicly available at \url{https://github.com/ababino/corona}.

\section*{Supplementary materials}

Materials and Methods

Supplementary Text

Figs. S1 to S5

Tables S1 to S4

References (31-35)

\bibliographystyle{Science}


\begin{figure}
	\centering
	\includegraphics[width=\linewidth]{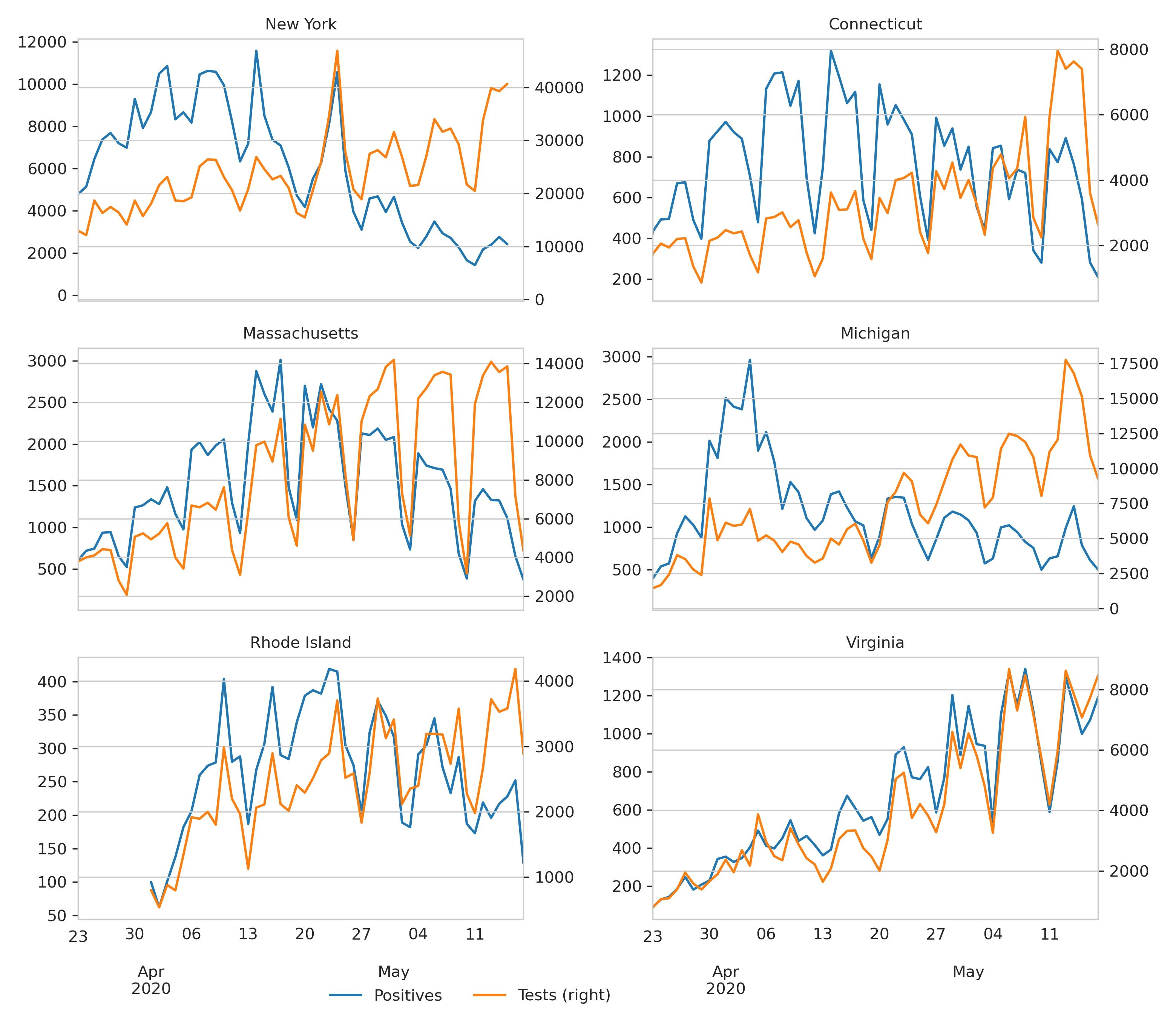}
	\caption{Daily number of new cases and test for each state in the dataset.}
	\label{fig:raw_data}
\end{figure}

\begin{figure}
	\centering
	\includegraphics[width=\linewidth]{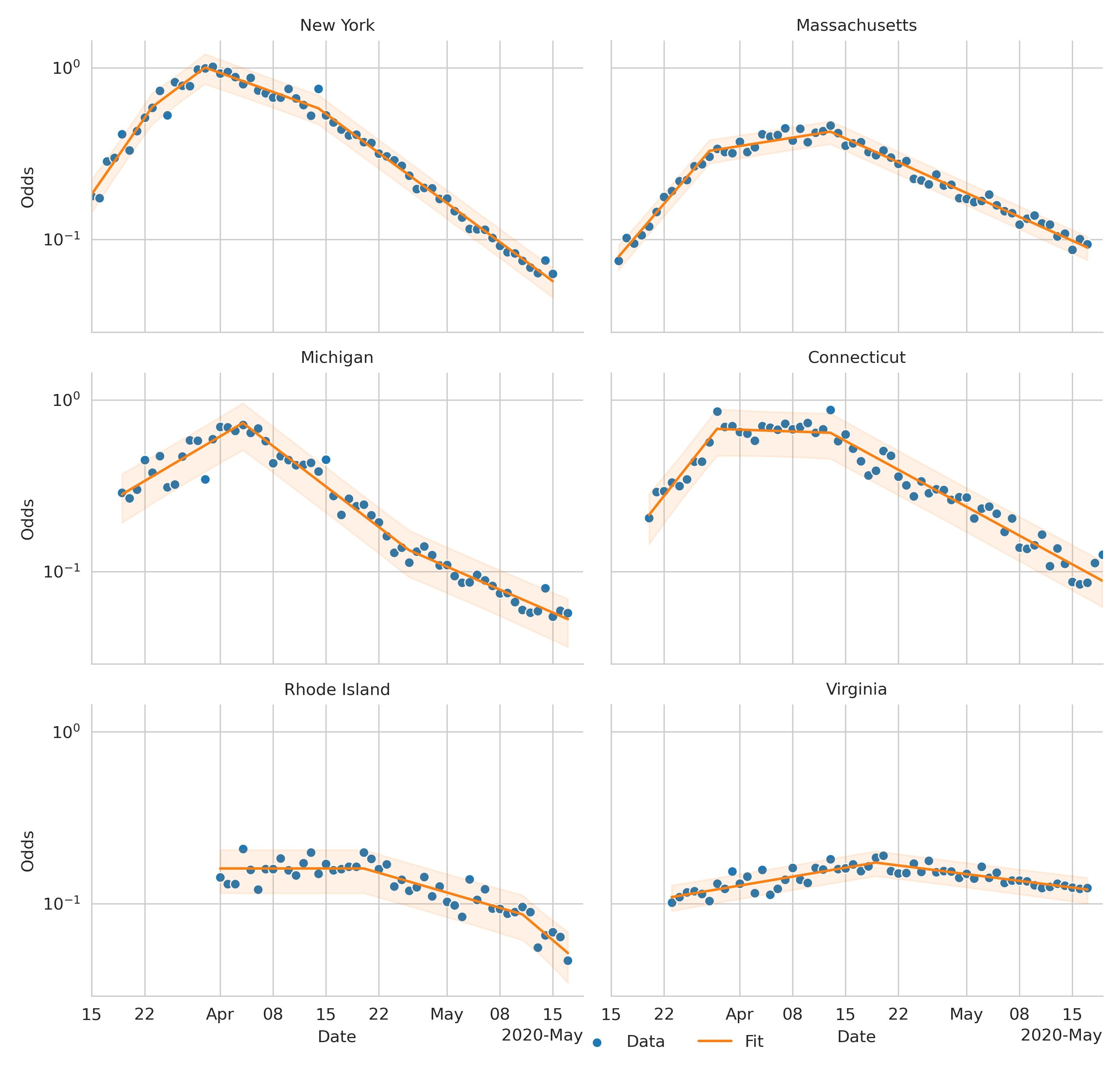}
	\caption{The odds of a positive test in logarithmic scale.
					 Under the assumptions of the model, this variable should be piece-wise linear.
					 The blue dots are the data points.
					 The orange lines regressed model and the orange shades are is the 95\% C.I.
					 }
	\label{fig:odds_fit}
\end{figure}

\begin{figure}
	\centering
	\includegraphics[width=\linewidth]{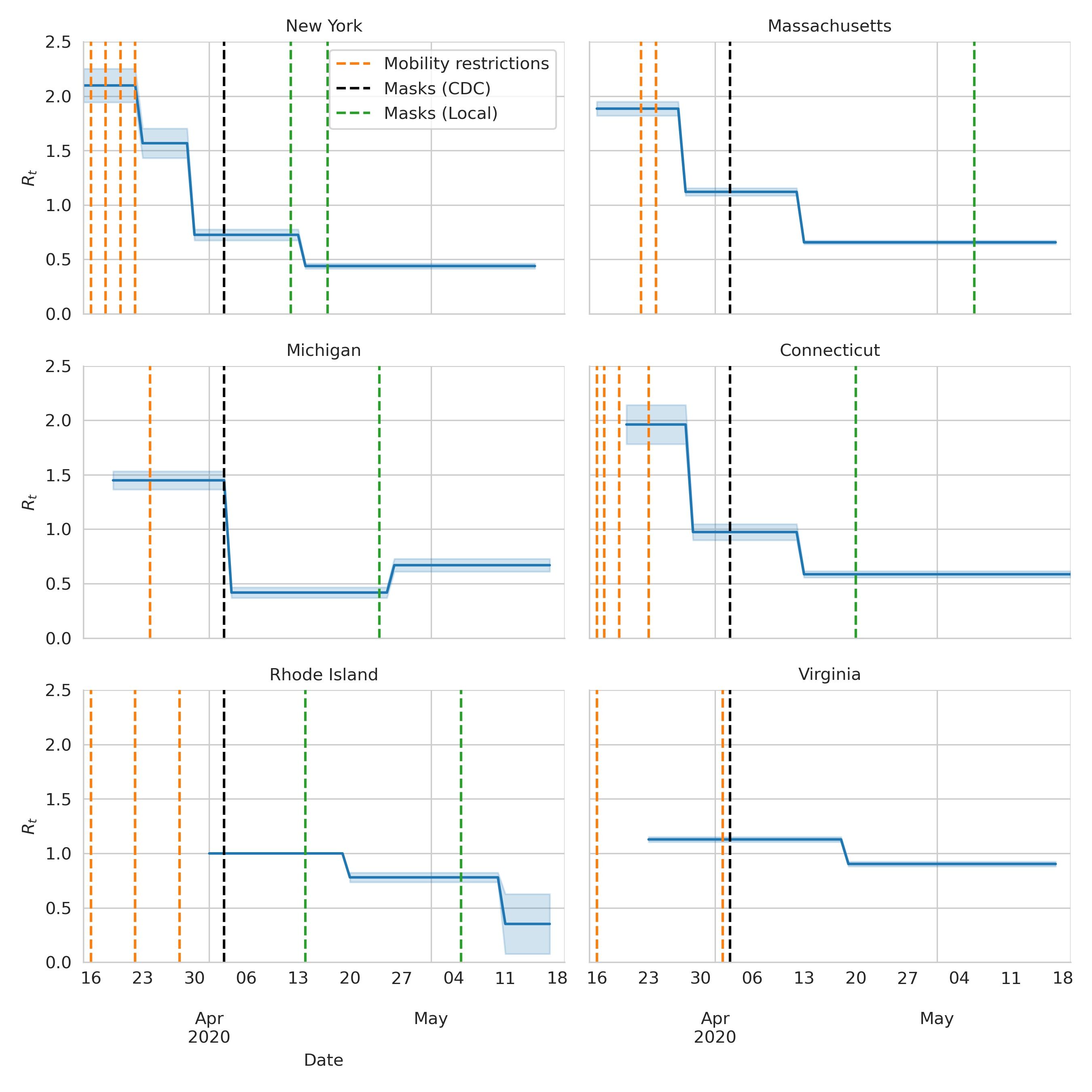}
	\caption{$R_t$ as a function of time.
					 The dashed vertical lines indicate different governmental interventions.
           Orange lines indicate mobility restriction orders such as closing bars, gyms, movie theaters, schools, and banning non essential work.
           The black line show the moment at which the CDC updated its guidelines to recommend wearing masks.
           The green lines show moments at which the local states changed their guidelines regarding masks.
           NY and RI enforced the use of masks for some jobs first, and later on they enforced mask wearing policies among the general public.
           MA and CT enforced the use masks by the general public.
           MI only enforced the use of masks in enclosed public areas, such as grocery stores.
           VA never enforced the use of masks.
           }
	\label{fig:rt}
\end{figure}

\begin{figure}
	\centering
	\includegraphics[width=0.75\linewidth]{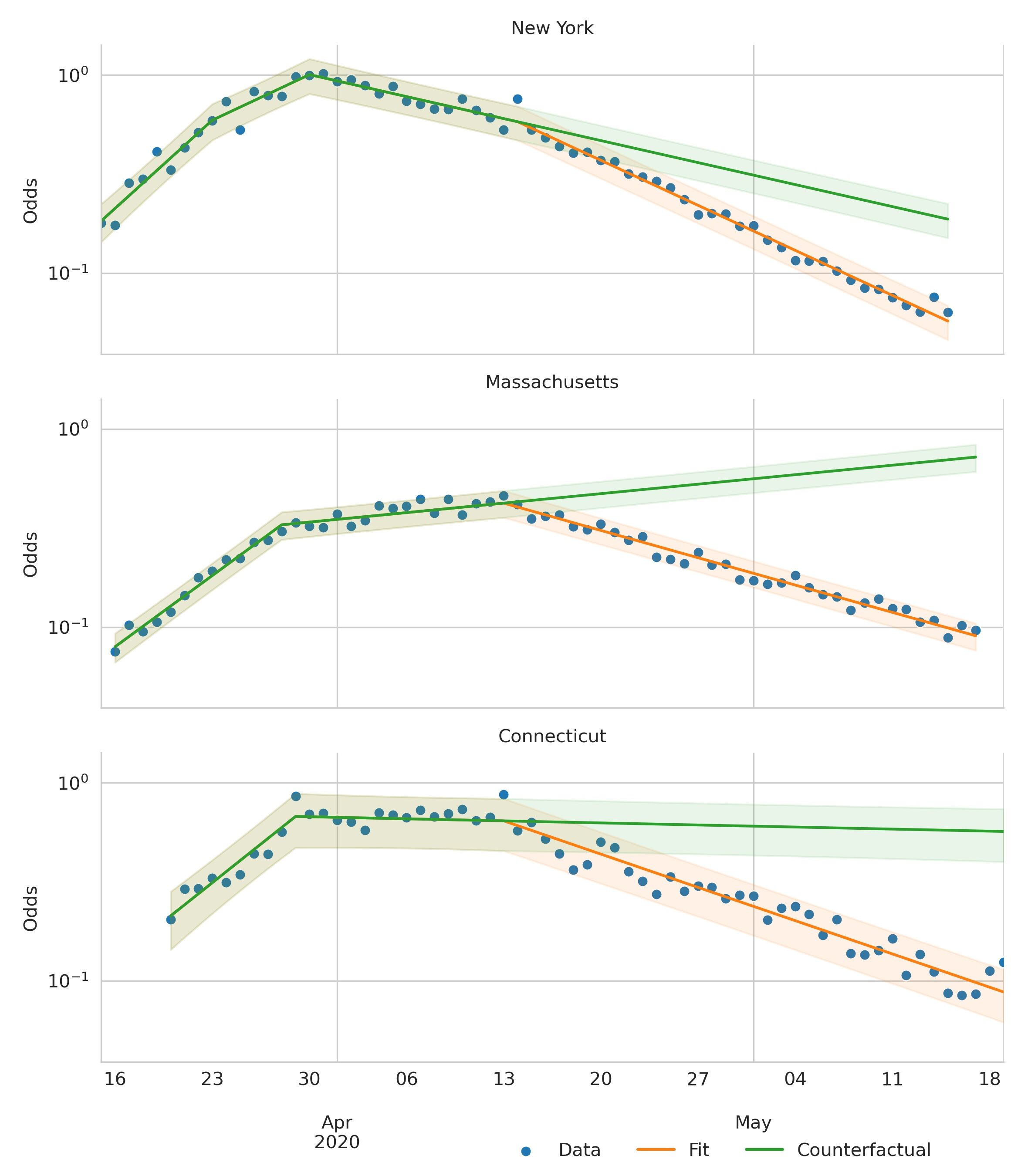}
	\caption{Actual and counterfactual scenarios.
	         The blue dots are the data points.
	         The orange lines show the regressed model, and the orange shades are the 95\% C.I.
	         The green lines show the counterfactual odds of the scenario where masks were not recommended; in causal inference jargon, \textit{do(not masks)}; and the green shaded areas are the 95\% C.I.
			}
	\label{fig:odds_cf}
\end{figure}

\section*{Supplementary Materials for\\Masks and COVID-19: a causal framework for imputing value to public-health interventions}
\noindent \textbf{This PDF file includes:}

	Materials and Methods

	Supplementary Text

	Figs. S1 to S5

	Tables S1 to S4

\section*{Materials and Methods}

\subsection*{Data}
We collected data from States that offer raw data on the number of tests and positive cases each day.
We found that 16 States offer that information.
Some of them offer an Abstraction Protocol Interface (API).
Others serve a file with the information.
Many states offer a visual dashboard with information on testing, but if they do not offer raw data, we did not use it.
There is at least one project that collects data on testing from all the states: \url{https://covidtracking.com/}.
This aggregator builds its database based on snapshots of the dashboards published by the states.
The information in these snapshots is averaged three or four times a day.
This process makes the accumulated number of tests and positives reliably, but not their daily change.
That is why we did not use information from this aggregator, and we only use direct information from official sources.
We provide links to each dataset in Table \ref{tab:datasets}

In the main text, we show results for six datasets with the highest $R^2$.
To show the robustness and also limitations of the framework here, we show an analysis for all the states where we found data on the daily number of cases and tests.
In the main text, we limit the analysis to the time in which NYS, MA, and CT ended their stay-at-home order.
To show the robustness of the framework, here we show the results when we apply the method to a bigger timespan.
Since, due to backlog, some states have a delay in reporting of about one week, we included data until the last day we find reliable data.

\subsection*{The odds as the dependent variable}
As can be seen in Fig. \ref{fig:raw_data} the number of daily positive tests, $P\! ositive_t$, oscillates in synchrony with the number of tests.
To overcome this source of noise, we propose to use the odds of a positive test:
$$
O\! dds_t = \frac{P\! ositive_t}{N\! egative_t}
$$
where $N\! egative_t$ is the number of negative tests on day $t$.

We show the number of positives and the number of tests for each dataset in Fig. \ref{fig:raw_data}

We show the evolution of the Odds in Fig. \ref{fig:odds_fit}.
The noise due to the variation in the number of tests is reduced, and a trend emerges.

\subsection*{The evolution of the Odds}
As shown previously (28), under the SIR model hypotheses, the number of newly infected individuals in a given day, $k_t$, can be approximated as:
$$
k_t = k_{t-1} e^{(R_{t-1}-1)\gamma}
$$

where $R_t$ is the instantaneous reproduction number (28), and $\gamma^{-1}$ is the average infectious period (26) estimated as \serialinterval\ days \serialintervalci\ according to (31) (in agreement with (32), but higher than reported in (33)).

Since we do not have access to the total number of infected individuals, but only to the tested population, we have to use some statistical assumptions about this population.
If we assume that the people being tested is a random sample of the population with \disease-like symptoms, we can state that\footnotemark \footnotetext{The following derivation was suggested to us by Will Meierjurgen Farr on this GitHub Issue \url{https://github.com/k-sys/covid-19/issues/45#issuecomment-623782130}}:

\begin{equation}
	P\! ositive_t = P_t(I|symptoms) P_t(symptoms) Nf_t
	\label{eq:positive_decomposition}
\end{equation}

where $P_t(I|symptoms)$ is the probability of a patient being positive for \virus\  given that she is symptomatic, $P_t(symptoms)$ is the probability of having \disease-like symptoms, $N$ is the total population, and $f_t$ is the fraction of people with symptoms that are selected to be tested (this number can be different each day, for example, if the number of tests available changes).
Similarly:
\begin{equation}
	N\! egative_t = P_t(not I|symptoms)P_t(symptoms)Nf_t
	\label{eq:negative_decomposition}
\end{equation}
where $P_t(not I|symptoms)$ is the probability of a patient being \virus\  negative given he has \disease-like symptoms.

Now, if we assume that $P_t(symptoms|I)=cte$, we can use Bayes theorem to show that:

$$
P_t(I|symptoms) P_t(symptoms) \propto P_t(I) = \frac{k_t}{N}
$$

Then:
\begin{equation}
P_t(I|symptoms) P_t(symptoms) \propto k_t
\label{eq:probas_prop_kt}
\end{equation}

Finally, if we assume that $P_t(not I|symptoms)P_t(symptoms)=cte$:
$$
O\! dds_t = \frac{P_t(I|symptoms) P_t(symptoms)Nf_t}{P_t(not I|symptoms)P_t(symptoms)Nf_t}
$$
$$
O\! dds_t = \frac{P_t(I|symptoms) P_t(symptoms)}{P_t(not I|symptoms)P_t(symptoms)}
$$
$$
O\! dds_t \propto k_t
$$

\begin{equation}
O\! dds_t = O\! dds_{t-1} e^{(R_{t-1}-1)\gamma}
\label{eq:odds_rt}
\end{equation}

We used four sets of hypotheses.
First, we use the assumptions of the SIR model.
Second, we use that the tested population is a random sample from the population with \disease-like symptoms (Eqs. \ref{eq:positive_decomposition} and \ref{eq:negative_decomposition}).
This assumption does not hold, for example, if the basis for testing someone is that she was in contact with a confirmed case.
Third, we assume that $P_t(not I|symptoms)P_t(symptoms)=cte$.
This hypothesis is equivalent to say that the number of people with \disease-like symptoms but without the \virus\  (for example, people with the flu) is constant, or its change rate is negligible compared with the change rate in the number of symptomatic people with \virus.
Fourth, we use that the symptoms show up instantaneously and that the tests are performed and processed on the same day (Eq. \ref{eq:probas_prop_kt}).
This last hypothesis is not true, and it is the reason why, in our analysis, the effects of the interventions show a delay to onset between 8 and 11 days.

\subsection*{Linearization}
We write Eq. \ref{eq:odds_rt} as a linear function of the rate of change of $R_t$.
Defining

\begin{equation}
	b_t =  e^{(R_{t}-1)\gamma}
	\label{eq:bi_rt}
\end{equation}

We can write Eq. \ref{eq:odds_rt} as:

\begin{equation}
O\! dds_t = b_{t-1} * O\! dds_{t-1}
\label{eq:odds1}
\end{equation}

Now, instead of using $b_t$ as the parameters to estimate we decompose each $b_t$ as follows:

\begin{equation}
	b_t = \prod_{i=0}^{t} a_i
	\label{eq:biai}
\end{equation}

The $a_i$s represent the rate of change of the variable $b_t$ in logarithmic scale.
Next, we replace the \ref{eq:biai} in \ref{eq:odds1}

$$
O\! dds_t = \prod_{i=1}^{t-1} a_i * O\! dds_{t-1}
$$
$$
O\! dds_t = \prod_{i=1}^{t-1} a_i * \prod_{i=1}^{t-2} a_i * O\! dds_{t-2}
$$
$$
O\! dds_t = \prod_{k=1}^{t-1}\prod_{i=1}^{k} a_i * O\! dds_{1}
$$
$$
O\! dds_t = \prod_{i=1}^{t-1} a_i^{t-i} * O\! dds_{1}
$$

Now, we linearize this result, and we generalize it to the case where $t=0$ using the $max$ function:

\begin{equation}
	log(O\! dds_t) = \sum_{i=1}^{max(t-1, 1)} (t-i)log(a_i)  +  log(O\! dds_{1})
	\label{eq:logodds}
\end{equation}

We can write \ref{eq:logodds} as a linear problem with the following definitions:

\begin{equation}
	y = X \beta + \beta_0
	\label{eq:linear_system}
\end{equation}

\begin{equation}
		y_t = log(O\! dds_t)
		\label{eq:y_log_odd}
\end{equation}
\begin{equation}
	X_{t,i} = max(t-i, 0)
	\label{eq:X}
\end{equation}
\begin{equation}
	\beta_t =  log(a_{t})
	\label{eq:beta_logai}
\end{equation}
Importantly, the SIR hypotheses are only necessary to draw the connection to $R_t$ (Eq. \ref{eq:bi_rt}).
However, Eq. \ref{eq:logodds} might hold even if the SIR hypotheses do not.
What would change is the interpretation of the parameters.

\subsection*{LASSO regression and feature selection}

Since in Eq. \ref{eq:linear_system}, we have as many regressors as samples, and we assume that the changes in $a$ are only due to top-down interventions we use a LASSO regression to fit the data (27).
This regression minimizes the loss function:
\begin{equation}
	Err = \frac{1}{n} \sum_{t=1}^n (y_t-\beta_0 - \sum_{i=1}^{n-1}\beta_i X_{t,i})^2 + \alpha \sum_{i=0}^{n-1}|\beta_i|
\end{equation}

This approach finds a sparse set of $\beta_i$.
We add two extra steps to sparsify even further this set of parameters.
If there are contiguous $\beta_i \ne 0$, we set to zero all of them but the first in the chunk.
Then, we fit the selected regressors using ordinary least squares, and we recursively remove the $beta_i$ with \mbox{$p$-$values^*>0.01$}, where \mbox{$p$-$values^*$} are the Bonferroni corrected \mbox{$p$-$values$}.
Using the LARS algorithm (34), we repeat these steps for different values of the hyperparameter $\alpha$, and we use the fit that minimizes the Bayesian Information Criterion (35).

We show the result of this procedure in Fig. \ref{fig:framework}.
In the top row of the figure, we show the change in the parameters as the value  of $\alpha$ changes.
In the second row, we show, for a given $\alpha$, the set of parameters yielded by the LASSO regression--most of these values are zero.
The dots in the second row are the parameters that are selected because they are the first of a non-zero chunk, and their \mbox{$p$-$values^*$} after applying a linear regression are smaller than $0.01$.
Finally, in the third row, we show the BIC value of all the possible models.
We show the goodness of fit of the final stage of the framework for each dataset in Table \ref{tab:gof}, and the value and statistics of each parameter in Table \ref{tab:coef}.

\subsection*{From fitted parameters to $R_t$}

To compute the value of $R_t$ from fitter parameters, we have to use equations \ref{eq:beta_logai}, \ref{eq:biai} and \ref{eq:bi_rt}.
From these equations, we arrive at the following equality:
\begin{equation}
  R_t = \frac{\sum_{i=0}^t \beta_i}{\gamma} + 1
\end{equation}
Where most of the $\beta_i$ values are zero.
Using this formula, we arrive at the values presented in Fig. 3 and Table 1 (main text).
We show the $R_t$ values for all states with data in Fig. \ref{fig:rt}.

\section*{Supplementary Text}
\subsection*{Simulations}

To test the framework, we show that it is able to estimate the parameters from simulated data.
We use the SIR equations:
\begin{equation}
\frac{dS}{dt} = -\beta \frac{SI}{N}
\end{equation}
\begin{equation}
\frac{dI}{dt} = \beta \frac{SI}{N} - \gamma I
\end{equation}
\begin{equation}
\frac{dR}{dt} = \gamma I
\end{equation}
where $S$ is the number of susceptible individuals, $I$ the number of infected individuals, $R$ the number of recovered (or death) individuals, $\beta$ is the contact rate and $\gamma$ the inverse of the average infectious period.
Also, $R_0=\beta/\gamma$

In our simulation we change the value of $\beta$ at times $t=21$, $t=41$, and $t=61$, from a initial value of $\beta=0.26$ to $\beta=0.2$, and then to $\beta=0.093$ and $\beta=0.053$.
Also we set $\gamma=7.5^{-1}$
We chose these values because the $R_t$ are similar to the one in the NYS dataset.
After carrying out the simulation we compute the \textit{log-odds}$=log(I/S)$ and add noise $\epsilon~N(0,0.05)$
We show the results applying the framework to this data in Fig. \ref{fig:SIR} and the real and estimated values in Table \ref{tab:SIR}.
Initially, the framework estimates $R_t=1.7$ $(1.95,2.0)$ $95\%$ CI.
This value is below the $R_0=2$ of the model, which is reasonable given that the framework should estimate the mean value of $R_t$ in a segment, which is always below $R_0$ (26).
The second discontinuity is found at $t=39$ and last one at $t=59$, two days before the real change.

This analysis shows that the framework can detect the number of discontinuities correctly, and it also estimates the value of $R_t$ accurately and the times of the breaks with no more than 2 days of error.

\begin{figure}
	\centering
	\includegraphics[width=\linewidth]{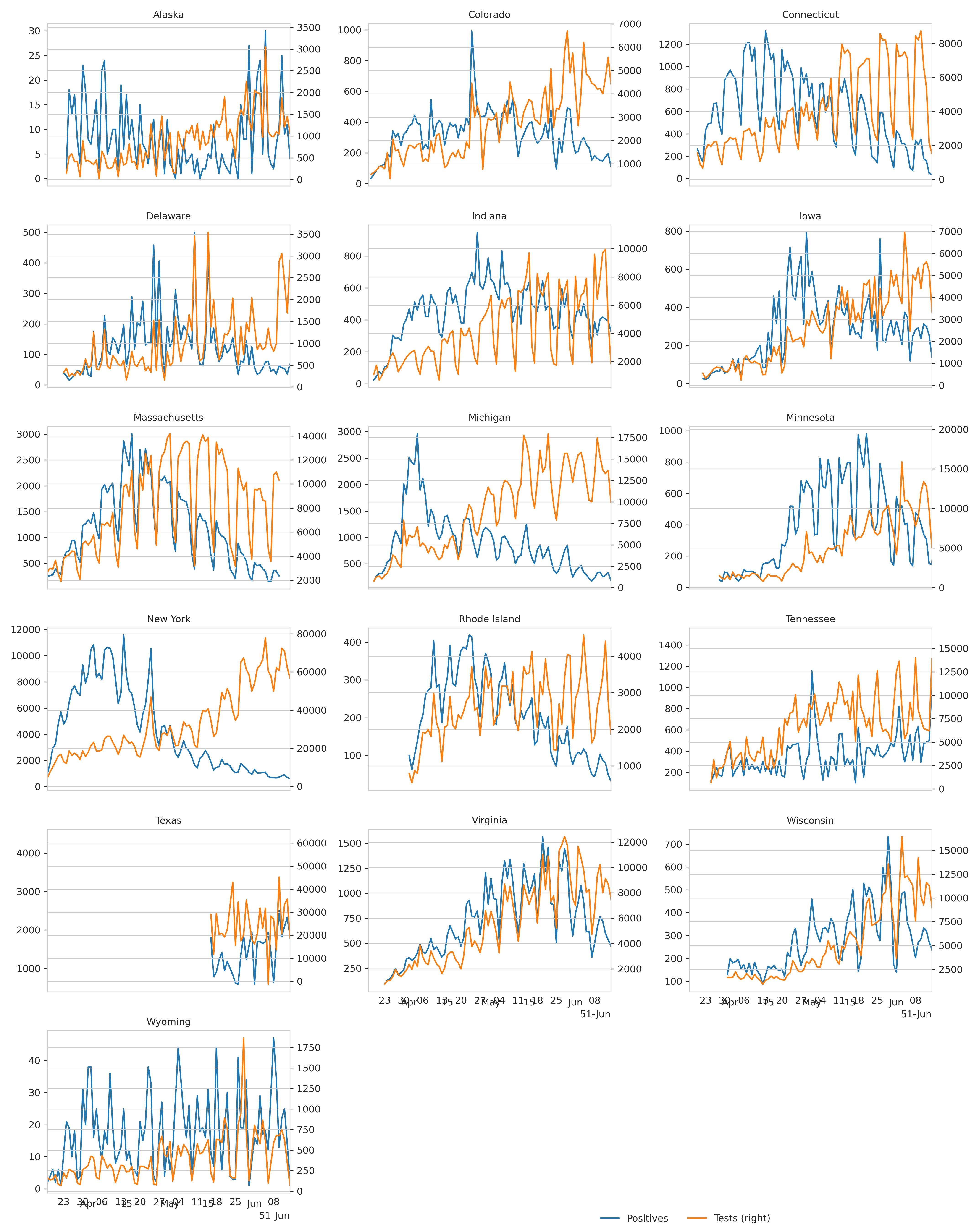}
	\caption{Daily number of new cases and tests for each state in the dataset.}
	\label{fig:raw_data}
\end{figure}

\begin{figure}
	\centering
	\includegraphics[width=\linewidth]{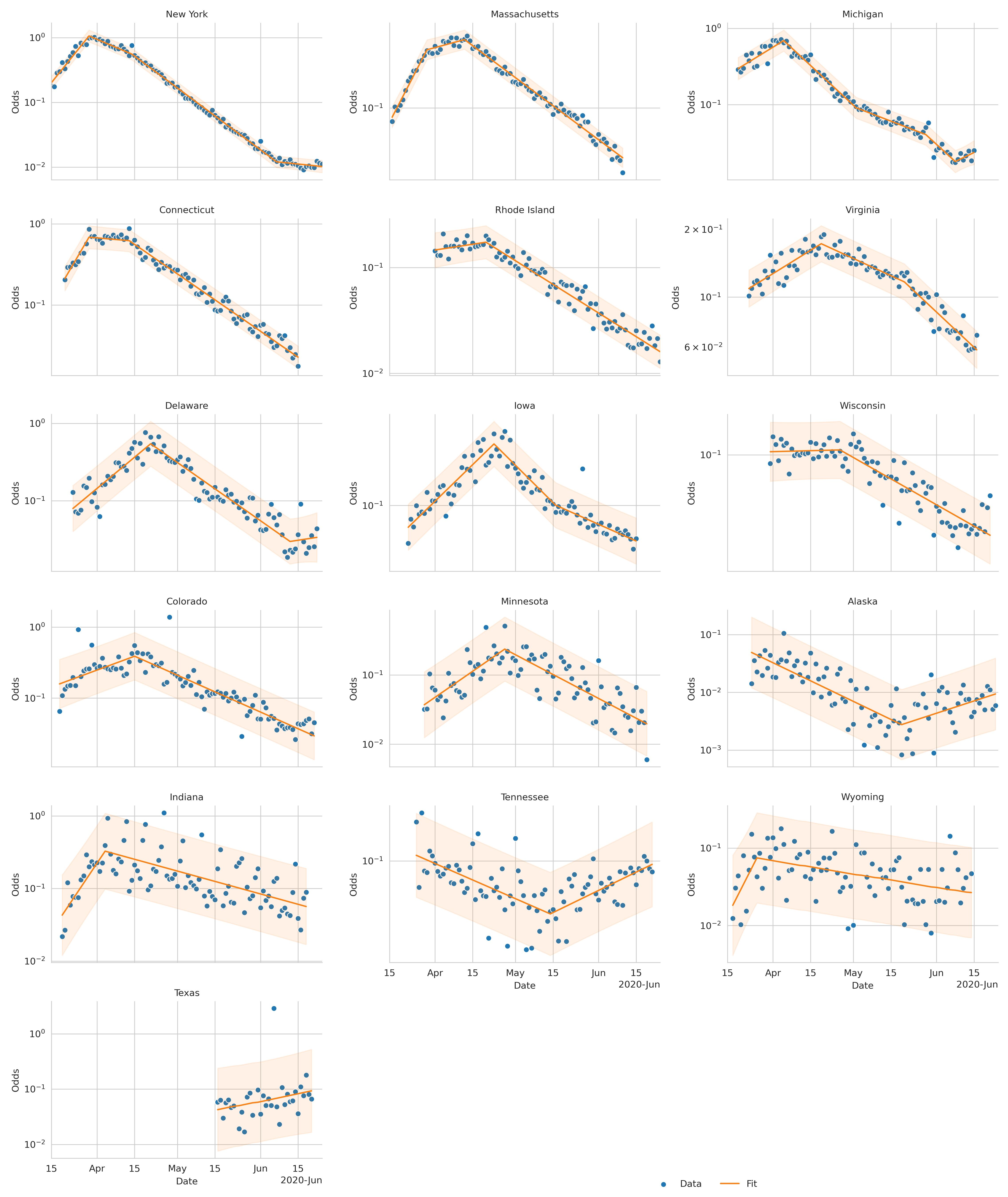}
	\caption{The odds of a positive test in logarithmic scale.
					 Under the assumptions of the model, this variable should be piecewise linear.
					 The blue dots are the data points.
					 The orange line is the LASSO fit, and the orange shade is the 95\% C.I.
					 }
	\label{fig:odds_fit}
\end{figure}

\begin{figure}
	\centering
	\includegraphics[width=\linewidth]{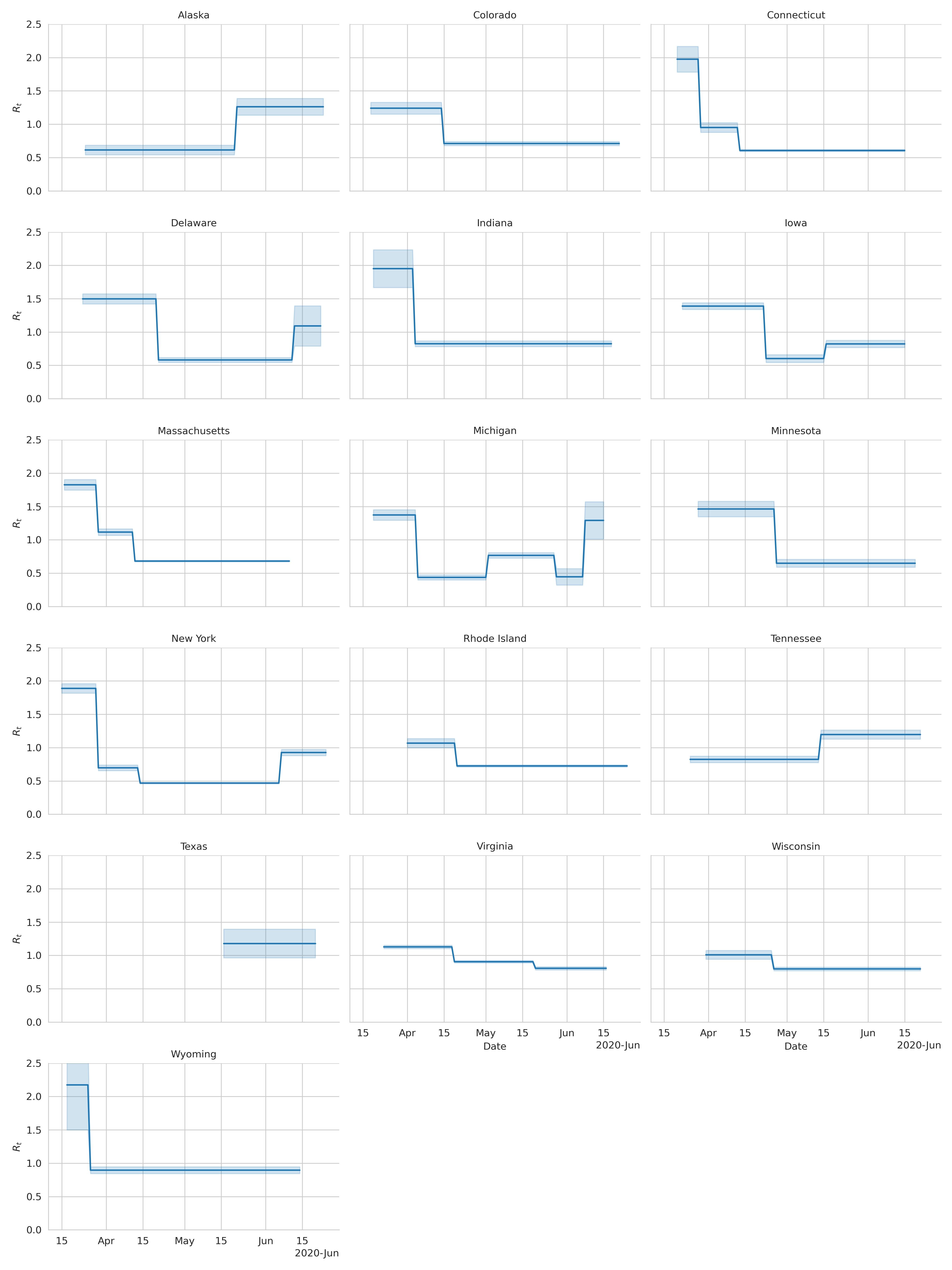}
	\caption{$R_t$ as a function of time.}
	\label{fig:rt}
\end{figure}

\begin{figure}
	\centering
	\includegraphics[width=\linewidth]{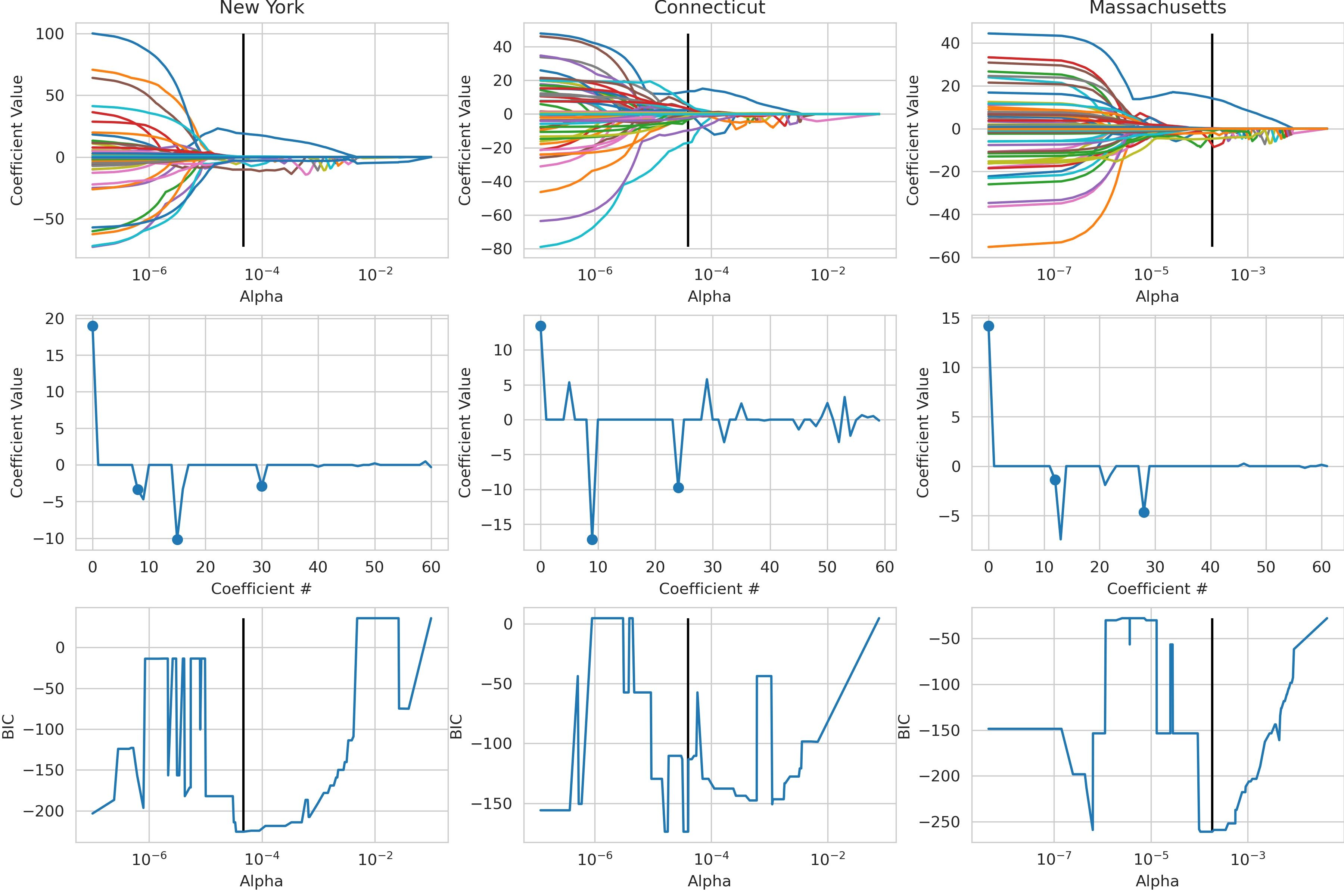}
	\caption{Steps of the framework.
           First row: coefficient values as a function of the parameter $\alpha$.
           As the value of $\alpha$ increases, the parameters collapse to zero.
           The black vertical line shows the selected value of $\alpha$.
           Second row: coefficient values at the selected $\alpha$.
           At the final $\alpha$, most of the coefficients are zero; if two of them are non zero in a row, we select the first one in the chunk.
           With the selected coefficients, we carry out a linear fit, and we iteratively remove the coefficients with \textit{p-value}$^*<0.01$.
           The dots indicate the parameters that were selected, either for being the first of a chunk or being significant.
           Third row: BIC as a function of $\alpha$.
           Since we performed these steps for all the $\alpha$ values, we compute the BIC for each of them.
           Then, we pick the one with the lowest BIC value.}
	\label{fig:framework}
\end{figure}

\begin{figure}
	\centering
	\includegraphics[width=\linewidth]{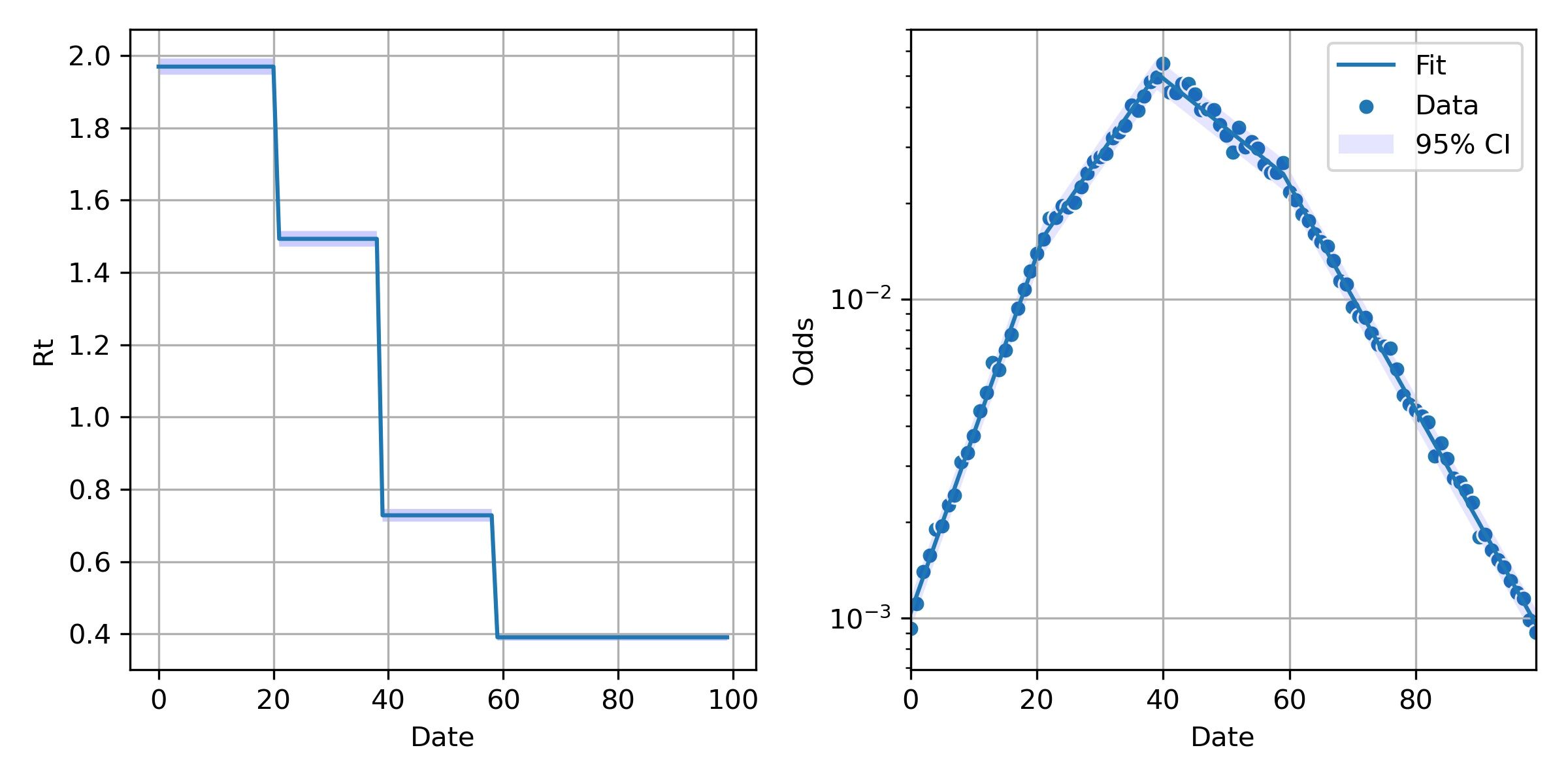}
	\caption{Simulated data using the SIR equations and the results from applying the framework.
          Left panel: estimated $R_t$ as a function of time.
          Right panel: simulated odds and final fit.}
	\label{fig:SIR}
\end{figure}

\begin{table}
\centering
\caption{Information about the source of the datasets used in this work}
\label{tab:datasets}
\begin{tabular}{p{2.3cm}p{12.5cm}}
	\hline

       Dataset &                                                                                                        Link to information  \\
\hline
        Alaska &                                https://coronavirus-response-alaska-dhss.hub.arcgis.com/datasets/daily-test-positivity/data  \\
      Colorado &                                  https://data-cdphe.opendata.arcgis.com/datasets/cdphe-covid19-daily-state-statistics/data  \\
   Connecticut &                 https://data.ct.gov/Health-and-Human-Services/COVID-19-PCR-Based-Test-Results-by-Date-of-Specime/qfkt-uahj  \\
      Delaware &                                                               https://myhealthycommunity.dhss.delaware.gov/locations/state  \\
       Indiana &                           https://hub.mph.in.gov/dataset/covid-19-case-trend/resource/182b6742-edac-442d-8eeb-62f96b17773e  \\
          Iowa &                                                                                              https://coronavirus.iowa.gov/  \\
 Massachusetts &                                                              https://www.mass.gov/info-details/covid-19-response-reporting  \\
      Michigan &                                                   https://www.michigan.gov/coronavirus/0,9753,7-406-98163\_98173---,00.html \\
     Minnesota &                                                         https://www.health.state.mn.us/diseases/coronavirus/situation.html  \\
      New York &                                      https://health.data.ny.gov/Health/New-York-State-Statewide-COVID-19-Testing/xdss-u53e  \\
  Rhode Island &                     https://docs.google.com/spreadsheets/d/1n-zMS9Al94CPj\_Tc3K7Adin-tN9x1RSjjx2UzJ4SV7Q/edit\#gid=590763272\\
     Tennessee &                                                       https://www.tn.gov/health/cedep/ncov/data/downloadable-datasets.html \\
         Texas &                                                                                    https://www.dshs.texas.gov/coronavirus/ \\
      Virginia &                                                                                  https://www.vdh.virginia.gov/coronavirus/ \\
     Wisconsin &                      https://data.dhsgis.wi.gov/datasets/covid-19-historical-data-table/data?where=GEO\%20\%3D\%20\%27State\%27 \\
       Wyoming &  https://health.wyo.gov/publichealth/infectious-disease-epidemiology-unit/disease/novel-coronavirus/covid-19-testing-data/ \\
			 \hline

\end{tabular}
\end{table}

\begin{table}
\centering
\caption{Goodness of fit for each dataset.}
\label{tab:gof}
\begin{tabular}{lrrrrr}
	\hline

         State & $R^2$ &    N &  D.F. &      F-value &        p-value \\
\hline
      New York & 0.996 &  102 &     4 &  6308.081088 &  2.989901e-116 \\
 Massachusetts & 0.984 &   87 &     3 &  1666.265804 &   5.057106e-74 \\
      Michigan & 0.983 &   89 &     5 &   987.866820 &   2.324393e-72 \\
   Connecticut & 0.981 &   88 &     3 &  1438.831575 &   4.534263e-72 \\
  Rhode Island & 0.952 &   85 &     2 &   810.885200 &   9.517308e-55 \\
      Virginia & 0.918 &   86 &     3 &   305.468584 &   2.184829e-44 \\
      Delaware & 0.886 &   92 &     3 &   228.607283 &   2.038413e-41 \\
          Iowa & 0.872 &   86 &     3 &   185.719315 &   1.862787e-36 \\
     Wisconsin & 0.811 &   83 &     2 &   171.824297 &   1.105016e-29 \\
      Colorado & 0.804 &   96 &     2 &   191.334575 &   1.095911e-33 \\
     Minnesota & 0.640 &   84 &     2 &    71.849515 &   1.131865e-18 \\
        Alaska & 0.589 &   89 &     2 &    61.705125 &   2.401719e-17 \\
       Indiana & 0.457 &   92 &     2 &    37.520871 &   1.521898e-12 \\
     Tennessee & 0.398 &   88 &     2 &    28.101998 &   4.283584e-10 \\
       Wyoming & 0.203 &   84 &     2 &    10.312295 &   1.023745e-04 \\
         Texas & 0.079 &   34 &     1 &     2.752898 &   1.068493e-01 \\
				 \hline

\end{tabular}
\end{table}

\begin{table}
\centering
\caption{Parameter values and statistics for the selected model for each dataset}
\label{tab:coef}
\begin{tabular}{llllr}
	\hline

       Dataset & Coef. Name & Coefficient &           95\% C.I. &  p-value \\
\hline
      New York &         x1 &     $0.146$ &    $(0.126, 0.167)$ & 1.94e-20 \\
      New York &         x2 &    $-0.071$ &  $(-0.106, -0.036)$ & 1.60e-04 \\
      New York &         x3 &    $-0.112$ &  $(-0.135, -0.089)$ & 7.51e-14 \\
      New York &         x4 &    $-0.038$ &  $(-0.047, -0.029)$ & 8.96e-12 \\
      New York &      const &     $-1.70$ &    $(-1.81, -1.59)$ & 4.50e-37 \\
   Connecticut &         x1 &     $0.128$ &    $(0.104, 0.152)$ & 2.90e-15 \\
   Connecticut &         x2 &    $-0.132$ &  $(-0.163, -0.101)$ & 9.93e-12 \\
   Connecticut &         x3 &    $-0.052$ &  $(-0.064, -0.039)$ & 2.57e-11 \\
   Connecticut &      const &     $-1.54$ &    $(-1.70, -1.39)$ & 6.68e-27 \\
 Massachusetts &         x1 &     $0.118$ &    $(0.110, 0.126)$ & 2.02e-38 \\
 Massachusetts &         x2 &    $-0.102$ &  $(-0.113, -0.091)$ & 2.55e-26 \\
 Massachusetts &         x3 &    $-0.062$ &  $(-0.068, -0.056)$ & 1.43e-28 \\
 Massachusetts &      const &     $-2.65$ &    $(-2.72, -2.58)$ & 5.36e-61 \\
      Michigan &         x1 &     $0.062$ &    $(0.052, 0.071)$ & 5.51e-19 \\
      Michigan &         x2 &    $-0.140$ &  $(-0.154, -0.126)$ & 2.03e-27 \\
      Michigan &         x3 &     $0.034$ &    $(0.022, 0.047)$ & 1.22e-06 \\
      Michigan &      const &     $-1.42$ &    $(-1.54, -1.30)$ & 1.67e-31 \\
  Rhode Island &         x1 &     $0.010$ &    $(0.001, 0.018)$ & 3.12e-02 \\
  Rhode Island &         x2 &    $-0.048$ &  $(-0.061, -0.035)$ & 2.45e-09 \\
  Rhode Island &      const &     $-1.93$ &    $(-2.04, -1.81)$ & 7.02e-33 \\
      Virginia &         x1 &    $0.0171$ &  $(0.0141, 0.0202)$ & 1.21e-15 \\
      Virginia &         x2 &    $-0.030$ &  $(-0.035, -0.025)$ & 1.18e-15 \\
      Virginia &      const &     $-2.22$ &    $(-2.27, -2.16)$ & 1.24e-56 \\
			\hline
\end{tabular}
\end{table}

\begin{table}
  \centering
	\caption{Simulated and estimated values of $R_t$}
  \begin{tabular}{ccccc}
		\hline
  $t$ & $\beta/\gamma$ & $\hat t$ &$R_t$ & $95\%$ $CI$ \\
  \hline
  $0$  & $2$   & $0$ & $1.97$ &	$(1.95,1.99)$\\
  $21$ & $1.5$ & $21$ & $1.50$ &	$(1.47,1.51)$\\
  $41$ & $0.7$ & $39$ & $0.73$ &	$(0.71,0.75)$\\
  $61$ & $0.4$ & $59$ & $0.39$ &	$(0.38,0.40)$\\
	\hline

  \end{tabular}
  \label{tab:SIR}
\end{table}

\end{document}